\documentstyle[aas2pp4,tighten]{article}

\begin{document}


\title{TWO-STREAM INSTABILITY OF COUNTER-ROTATING GALAXIES}

\author{R.V.E. Lovelace, K.P. Jore, and M.P. Haynes}

\affil{Department of Astronomy, Cornell University,
Ithaca, NY  14853-6801}

\begin{abstract}

The present study of the two-stream instability in stellar disks
with counter-rotating components of stars and/or gas is
stimulated by recently discovered counter-rotating
spiral and S0 galaxies.  Strong linear two-stream instability
of tightly-wrapped spiral waves is found for one and two-armed
waves with the pattern angular speed of the unstable waves
always intermediate between the angular speed of the
co-rotating matter ($+\Omega$) and that of the counter-rotating
matter ($-\Omega$).  The instability arises from the interaction
of positive and negative energy modes in the co- and counter-rotating
components.  The unstable waves are in general convective - they
move in radius and radial wavenumber space - with
the result that amplification of the
advected wave is more important than the local
growth rate.  For a galaxy of co-rotating stars and counter-rotating
stars of mass-fraction $\xi_* < {1\over 2}$, or of counter-rotating
gas of mass-fraction $\xi_g < {1\over 2}$, the largest
amplification is usually for the one-armed leading waves
(with respect to the co-rotating stars).
For the case of both counter-rotating stars and gas,
the largest amplifications are for $\xi_*+\xi_g \approx
{1\over 2}$, also for one-armed leading waves.  The two-armed
trailing waves usually have smaller amplifications.
The growth rates and amplifications all decrease
as the velocity spreads of the stars and/or gas increase.  It
is suggested that the spiral waves can provide an effective
viscosity for the gas causing its accretion.

\end{abstract}


\section {INTRODUCTION}

A surprise of recent high spectral resolution studies of normal galaxies
is the occurrence of counter-rotating gas and/or stars in galaxies
of all morphological types, ellipticals, spirals, and irregulars (see
review by Galletta 1996).  In the ellipticals, the counter-rotating
component is usually in the nuclear core and may result from merging
of galaxies with opposite angular momentum.  In contrast, in a number
of spiral and $S0$ galaxies, counter-rotating disks of gas and/or stars
have been found to co-exist with the primary
disk out to large distances ($10-20$ kpc).
Examples include NGC 4550 (Rubin, Graham, \& Kenney 1992; Rix et al. 1992),
NGC 4826 (Braun et al. 1992), NGC 7217 (Merrifield \& Kuijken 1994),
NGC 4546 (Sage \& Galetta 1994),
NGC 3626 (Ciri et al. 1995), NGC 3593 (Bertola et al. 1996), and
NGC 4138 (Jore, Broeils, \& Haynes 1996).  Table 1 summarizes data
on these galaxies.  These galaxies are early type and have ring(s) or
dust lanes or other morphological peculiarities.   Related
cases include galaxies with infalling HI gas, for example, NGC 4254 which has
an $m=1$ spiral arm (Phookun, Vogel, \& Mundy 1993), and NGC 628 which has a
warped and distorted outer disk with high-velocity
clouds (Kamphius \& Briggs 1992).

It is not likely that
the large-scale counter-rotating components result from mergers
of similar mass galaxies of opposite angular momenta because of the
large vertical thickening observed in simulation studies of such
mergers (Barnes 1992).  Formation of disk galaxies by accretion
has been studied by Ryden and Gunn (1987) and Ryden (1988).
Quinn and Binney (1992) have shown
that the more recently accreted matter at
large distances has a spin anticorrelated with
the spin of the central mattter.
Thakar and Ryden (1996) discuss different possibilities for the formation
of counter-rotating galaxies, (1) that the
counter-rotating matter may come from the merger of an oppositely
rotating gas rich dwarf
galaxy with an existing spiral, and (2) that the accretion of gas
may occur over the lifetime of a galaxy with the more recently
accreted gas counter-rotating.

An important open theoretical question is: What is the interaction
between the co- and counter-rotating components observed on large-scales
in spiral and $S0$ galaxies, and does this interaction facilitate accretion
of the counter-rotating component?  The component without neutral or ionized
gas rotating in the same direction is referred to as the
co-rotating component.  Clearly, a strong interaction  will
occur between co- and counter-rotating gas with the result that these
components are likely to be spatially separated (see however Lovelace
and Chou 1996).  In this work we assume that any co-rotating
gas has been swept-out from
the spatial regions (radii) occupied by counter-rotating gas.
The interaction
between co- and counter-rotating stars by dynamical friction is negligible
because of the small densities and large relative velocities.  Also, the
direct interaction of co-rotating stars with counter-rotating gas is
negligible because of the small cross-section of a star.
Accretion occurs in the counter-rotating galaxy model of Thakar and
Ryden (1996)
due to viscosity of the gas which is treated using sticky-particle $N$-body
simulations.  However, the viscosity used is not derived from
physical principles.   Here, we
analyze the influence of the two-stream instability in generating
spiral waves in counter-steaming flat galaxies and suggest that these
waves may give rise to an effective viscosity for counter-rotating gas.

Theoretical interest in galaxies with both co- and counter-rotating
stars was stimulated by the early stability analysis of Kalnajs (1977) who
found that the counter-rotating stars have a stabilizing influence
on the bar forming ($m=2$) mode in models without a massive halo.
Studies of counter-streaming stability of
idealized self-gravitating systems were made
by a number of authors, by Bisnovatyi-Kogan
et al. (1969) and Bisnovatyi-Kogan (1973) for counter-rotating stars
in a finite radius cylinder, by Araki (1987) for equal populations of
counter-rotating stars in finite-radius rigidly rotating
Kalnajs disks, and by Lynden-Bell (1967)
and Araki (1987) for a homogeneous system of counter-streaming stars
following Jeans' neglect of the background potential.  Perturbations
of the form $f(r){\exp}(im\phi-i\omega t)$ (in cylindrical coordinates
$(r, \phi, z)$ with $m=0,1,..$) were found to be linearly
unstable for some conditions for the cylinder for $m=2$, whereas
for the Kalnajs disks the $m=1$
mode appears as the main two-stream instability.
For homogeneous counter-streaming systems with Maxwellian
distribution functions, Lynden-Bell and Araki found that
the Jeans instability always dominates the two-stream instability.
N-body computer simulations of stellar disks by Zang and Hohl (1978),
Sellwood and Merrit (1994), and Howard et al. (1996) support
the idea of Kalnajs (1977) that counter-rotating stars have
a stabilizing influence on the bar forming mode.  However, the
counter-streaming is found to enhance the growth of the one-armed
($m=1$) mode.

Section 2 of this work discuss the basic equations for the small-amplitude,
tightly-wrapped spiral waves in a flat galaxy.
Section 3 treats the stability of the spiral waves
for different cases of a disk with co-rotating stars and counter-rotating
gas and/or stars.
Section 4
shows that the two-stream instability found in Sec. 3
arises from the interaction of positive
and negative energy modes associated with the co- and counter-rotating
matter.  Section 4 discusses possible non-linear effects
of the instabilities including gas accretion.
Section 5 gives conclusions of this work.

\section {THEORY OF SPIRAL WAVES}

We first
give a brief summary of the linear WKB
theory of tightly-wrapped spiral waves in a
single component galaxy of stars rotating with
angular rate $\Omega(r) >0$ (see for example
Binney \& Tremaine 1987; hereafter denoted
BT; or Palmer 1994).
We use an inertial, cylindrical
$(r,\phi,z)$ coordinate system and assume
a thin disk galaxy as indicated in Figure 1.
 In the midplane of
the galaxy the perturbation of the gravitational
potential is
$$ \delta \Phi(r,\phi,z=0,t) =
C~\exp\big[~i\int^r dr'~k_r(r') +im\phi -i\omega t \big],
\eqno(1)$$
where the radial wavenumber $k_r$ satisfies
$|k_r r|\gg1$ for a tightly wrapped wave, $C(r)$ a slowly
varying function of $r$, $m=\pm1,\pm2,..$
is the number of spiral arms, and $\omega$ is the angular frequency of the
wave.
Only $m>0$
need be considered because $\delta \Phi^*$ is
a valid solution if $\delta \Phi$ is.  Then, $k_r>0$ ($<0$) corresponds
to a trailing (leading) spiral wave.  The thin disk assumption requires
$|k_r h| \ll 1$,
where $h$ is disk half-thickness.

A well-known
calculation (see BT or Palmer 1994) of the linear
perturbation of the dynamical equations leads to the dispersion relation
$$0 = \epsilon(\omega, k_r) \equiv 1+ {\cal P}_*(\omega,k_r)~,
\eqno(2)$$
where
$${\cal P}_* = {{2|\overline k_r|\exp(-X_*)}\over X_*}
\sum_{l=1,2,..}{{I_l(X_*)}\over (s/l)^2-1}~,
$$
$$ s \equiv (\omega-m\Omega)/\kappa~,~~~~~~\kappa^2 \equiv
{{1}\over{r^3}}{d \over {dr}}
(r^4 \Omega^2)~,$$
$$\overline k_r \equiv k_r/k_{crit}~,~~~~~~k_{crit} \equiv {{\kappa^2}\over
{2\pi G \Sigma_{tot}}}~, $$
$$X_*\equiv \bigg({{k_r \sigma_r}\over {\kappa}}\bigg)^2 = 0.28568 (Q_*
\overline k_r)^2~,$$
$$Q_* \equiv {{\kappa \sigma_r}\over {3.3583 G\Sigma_{tot}}}~.$$
Here, $\epsilon(\omega, k_r)$ has the role of a
dielectric function for the disk, and $\cal{P}_*$ the
polarization function (the $*$-subscript denotes stars);
$s$ is a dimensionless frequency;  $\kappa(r)$
is the radial epicyclic frequency of a star;  $k_{crit}$
is a characteristic wavenumber, and $\overline
k_r$ is the dimensionless radial wavenumber;
$\Sigma_{tot}(r)$ is the total surface mass density
of the disk;  $Q_*(r)$ is Toomre's stability
parameter;
$\sigma_r(r)$ is the radial dispersion of the star
velocities (for a stellar distribution function
$f_* \propto \exp[-v_r^2/(2\sigma_r^2)])$;
and $I_l$ is the usual modified Bessel function of
order $l$.  The condition for
axisymmetric ($m=0$) stability is $Q_*>1$ (Toomre 1964).
The influence of halo matter enters through $\Omega(r)$.

Figure 2 shows wavenumber ($k_r$) - frequency ($s$)
plots for two values of $Q_*$ obtained from equation (2).
The corresponding
plot in BT (Figure 6-14a) and in Palmer (1994, Figure 12.2) is
{\it incomplete} because the higher order
branches, labeled $l=2, 3,..$, have been omitted.  Further,
the statement in BT that ``a stellar disk has no pressure
forces and therefore cannot support waves with $|s|>1$'' (page 367)
is incorrect.  The branches $l=2,3,..$ are the analogues of the
well-known Bernstein modes which propagate across a uniform magnetic
field in a collisionless plasma (see for example
Krall \& Trivelpiece 1973).  Here, the
coriolis force is the analogue of the
Lorentz force in a magnetized plasma.  Note in particular that the
Linblad
resonances correspond to $s=\pm l$ or $\Omega_p = \Omega \pm l\kappa/m$,
where $\Omega_p \equiv \omega/m$ is the pattern angular speed, $m=1,2,..$,
{\it and}
$\l = 1,2,..$.

For later use, we also give the dispersion relation for a single
component gaseous disk with rotation rate $\Omega(r)>0$,
$$0=\epsilon(\omega,k_r) \equiv 1 + {\cal P}_g(\omega,k_r)~,
\eqno(3)$$
where
$$ {\cal P}_g = {{|\overline k_r|}\over {s^2-1-(k_r c_s/\kappa)^2}}~,$$
where $c_s$ is the sound speed in the gas, and where the other quantities
are the same as in equation (2).   Equation (3) gives the explicit
dependence $s^2=1-|\overline k_r|+(k_r c_s/\kappa)^2$.  Here, the
Toomre (1964)
condition for axisymmetric stability is $c_s > \kappa/(2 k_{crit})$ or
$Q_g \equiv \kappa c_s/(\pi G \Sigma_{tot}) >1$.

Note that $\Omega^2$, $\kappa^2$, and $\kappa \equiv |\kappa|$ are the
same for both co- and counter-rotating disk components.  Note also that
$\Sigma_{tot}$ is the total - gas plus stellar - mass density in all
components.  In that ${\cal P}_*$ and ${\cal P}_g$ are even functions
of $k_r$, we have $\omega(k_r) = \omega(-k_r)$.

\medskip

\section{TWO-STREAM INSTABILITY}

In the following subsections we consider the two-stream instability
in a number of different limits.  Specifically, we consider co-rotating
stars ($+\Omega$) and in subsection

 3.1~~counter-rotating gas ($-\Omega$),

 3.2~~a small mass-fraction of counter-rotating gas,

 3.3~~a large mass-fraction of
counter-rotating gas,

 3.4~~a small mass-fraction of counter-rotating stars,

 3.5~~a large mass-fraction of counter-rotating stars, and

 3.6~~arbitrary fractions of counter-rotating gas and stars.

\subsection{Co-Rotating Stars/Counter-Rotating Gas}

The dispersion relation for a two component galaxy consisting of
co-rotating stars ($+\Omega$) and counter-rotating gas ($-\Omega$)
can be written down immediately using equations (2) and (3),
$$0 = \epsilon(\omega,k_r) \equiv 1+(1-\xi_g){\cal P}_*(s,k_r)+
\xi_g{\cal P}_g(s+w,k_r)~,~
\eqno(4)$$
where $w \equiv {2m\Omega}/{\kappa}~,$  and
where $$\xi_g \equiv \Sigma_g/\Sigma_{tot}$$ is the
fraction of the disk surface mass density in gas.
To understand equation (4) it is useful to
introduce the notion of {\it stellar modes}
which obey
$$ 0 = 1+(1-\xi_g){\cal P}_*(s,k_r)~, \eqno(5a)$$
and {\it gas modes} which obey
$$0 = 1 +\xi_g{\cal P}_g(s+w,k_r)~.\eqno(5b)$$
Equation (5a) gives a family of curves $s=s_*(k_r)$ for the star modes,
while (5b) gives curves $s=s_g(k_r)$ for the gas modes.  This is shown
in Figure 3 for $m=2$.

The approximation of equations (5) breaks down near the isolated points in
the $(k_r,s)$ plane where the star and gas modes
cross, $s_*(k_r) = s_g(k_r)$.
At these points there is a strong resonant interaction between the modes.

\subsection{Co-Rotating Stars/Low-Mass Counter-Rotating Gas}

For $\xi_g \ll 1$, an approximate solution of the dispersion
relation (4) near a mode-crossing point can easily be obtained as
follows.
To zeroth order in $\xi_g$, equation (4) is satisfied by $(k_r,s_o)$
obeying $0 = 1 + {\cal P}_*(s_o,k_r)$.  For small $\xi_g$,  the
gas response is large if
$\varepsilon \equiv [(s_o+w)^2 - 1 - (k_r c_s/\kappa)^2]/[2(s_o+w)]$ is
small.
To first order in $\xi_g$, the
solution of equation (4) can thus be written as $s = s_o+\delta s$ with
$|\delta s| \ll |s_o|$.  Taylor expanding about $s_o$ gives
$$\delta s = {{\delta \omega}\over{\kappa}} =
{-\varepsilon} \pm \big( {\varepsilon}^2
+\delta s_o^2\big)^{1\over 2},
\eqno(6)$$
where
$$\delta s^2_o \equiv
 {{-~\xi_g ~|\overline k_r|}\over{2(s_o+w)(\partial {\cal P}_*/{\partial
s_o})}}~.
$$
Instability, $\omega_i \equiv Im(\omega) >0$, occurs if $\delta s_o^2 <0$ and
$\varepsilon^2 < |\delta s_o^2|$.  (The damping roots with $\omega_i <0$ are
ignored here and subsequently.) From equation (2), 
$\partial{\cal P}_*/{\partial s_o} = - s_o|..|$.
Thus for instability $s_o(s_o+w)<0$.
Equivalently, there is instability if the frequency of the
mode crossing, $\omega_o = \kappa s_o + m\Omega$, obeys
$$ -\Omega < ~{{\omega_o}\over{m}}~ < \Omega~,
\eqno(7)$$
where $\omega_o/m$ is the pattern velocity of the perturbation.  Later,
in Sec. 4, we give a more general treatment of instability
in counter rotating galaxies which also leads to equation (7).  From
equation (6), note that the maximum growth rate, $max(\omega_i) =  \kappa
|\delta s_o|$,
occurs for $\varepsilon = 0$ and scales
as $\xi_g^{1\over 2}$.

Figure 4 shows the behavior of the mode crossings for an unstable and
a stable crossing of Figure 3 obtained from equation (4).  For $\xi_g \ll 1$
and $c_s/\sigma_r = 1$, there are no unstable crossings for $m=1$,
while for $m=3$ there are unstable crossings near $\overline k_r \approx
1.04,~2.9,$ and $4.24$.  In all cases the unstable crossings obey
equation (7).

Consider the consequence of the wave growth.  For this
it is useful to examine the evolution of a wave packet.
The centroid of the packet is at $r$ and at $k_r$ in wavenumber space.
The dispersion relation (4) gives $\omega_r =
\omega_r(r, k_r) =$  const. which has the role of a Hamiltonian
for the packet.  The influence of the imaginary part of $\omega$ is
discussed later. The Hamiltonian equations are
$$ {{dr}\over {dt}} = v_g = {{\partial \omega_r}\over {\partial k_r}}~,~~~~~~
{{dk_r}\over {dt}} = -~{{\partial \omega_r}\over {\partial r}}~,
\eqno(8)$$
where $v_g$ is the group velocity (see for example Bekefi 1966). In general,
the $r$ dependence of $\omega_r$ results from the $r-$dependence of
all of the quantities, $\Omega, ~\kappa, ~k_{crit},$ etc., which enter in
equation (4).
To make the discussion tractable we assume that
$\Omega/\kappa,~Q_*,~c_s/\sigma_r,$
and $\xi_g$ are independent of $r$.  Then we have $\omega_r = \Omega(r)
f(\overline k_r)$,
where $\overline k_r \equiv k_r/k_{crit}(r)$ and where $f$ is a
dimensionless function.
Notice that Figure 4 gives $f = \omega_r/\Omega$ as a function of
$\overline k_r$. From equations (8), we find
$$ {{dr}\over{dt}} = {{\Omega}\over {k_{crit}}}~{{\partial f}\over
{\partial \overline k_r}}~,~~~~~~
{{d \overline k_r}\over {dt}} = -~{{\omega_r}\over
{k_{crit}\Omega}}~{{\partial \Omega}\over
{\partial r}}~.
\eqno(9)$$
For a flat rotation curve galaxy, $\Omega = \Omega_o (r_o/r)$ with
$\Omega_o, r_o =$ constants,
we find $d \overline k_r/dt = \omega_r/(k_{crit}r)$, and
$r = r_o(\Omega_o/\omega_r)f(\overline k_r)$.
We may take $r_o$ to be the initial radius of the wave packet.  Therefore,
for the
case of Figure 4a, we see that the packet moves radially outward in the
unstable range
of $\overline k_r$ and that $\overline k_r$ increases monotonically in that
$\omega_r > 0$.
It is a trailing spiral wave.
(For $\overline k_r < 0$, $f(\overline k_r)$, which is an even
function of $\overline k_r$, decreases with $\overline k_r$ in the
corresponding unstable region and the packet
moves inward while $\overline k_r$ increases.  The wave
in this case is a leading spiral.)

The maximum amplification of a wave packet results when $\overline k_r$
increases through all of the unstable range of $\overline k_r$ in which
$f(\overline k_r) > 0$ (or $<0$).
This range is denoted $\Delta \overline k_r$.
This gives an amplification factor
$$A \equiv \exp\big(\int dt~ \omega_i \big)
= \exp\bigg( {{\Omega_or_o }\over { \omega_r}}
\int_{\Delta \overline k_r} d\overline k_r~ k_{crit}
{{\omega_i(\overline k_r)}\over{\Omega}}\bigg)~.
\eqno(10)$$
In general, $A$ will be
less
than $ A _{max} =\exp[2\pi (T_{gal}/$ $T_{rot})(\omega_i/\Omega)_{max}]$,
where $T_{gal}$ is the
age of the galaxy, $T_{rot}$ is the rotation period at the
considered location ($r$), and $(\omega_i/\Omega)_{max}$ is the maximum
growth rate (as a function of $\overline k_r$).

For the case $m=2$ of Figure 4a,
the argument of the exponential is about $0.01(k_{crit} r_o)$.
For values of $k_{crit}$ of the order of that estimated for
our galaxy
$ \approx 2\pi/10kpc$ at $r=8.5kpc$ (BT), we conclude that the wave
growth is insignificant.  Basically, the perturbation
convects out of the unstable $\overline k_r$ range before
there is appreciable growth.  The same conclusion applies to the unstable
$m=3$ mode crossings for $\xi_g \ll 1$ and $c_s/\sigma_r = 1$.
\medskip

\subsection{Co-Rotating Stars/High-Mass Counter-Rotating Gas}

If the mass fraction of gas $~\xi_g~$ is not small
compared with unity, the axisymmetric stability
of the disk is affected by both the sound speed in the gas ($c_s$) and
the radial velocity spread of the stars ($\sigma_r$ or $Q_*$).  For
$m=0$, equation (4) is a function of $\omega^2$, and the
stability/instability boundary $\omega^2 = 0$ gives
$$0 = 1-{{(1-\xi_g)|\overline
k_r|}\over{X_*}}\bigg(1-\exp(-X_*)\big)I_o(X_*)\bigg)
$$ $$-{{\xi_g|\overline k_r|}\over{1+(k_r c_s/\kappa)^2}}~,
\eqno(11)$$
where $X_* = (k_r\sigma_r/\kappa)^2$ (Toomre 1964).
For a given $\sigma_r$ and $\xi_g$, the absence of solutions of
equation (9) for $k_r$ for a critical, sufficiently
large $c_s$ implies axisymmetric stability.
This 
is shown in Figure 5.  In contrast with the propagating or {\it convective}
instability 
discussed above, the $m=0$ instability gives $\omega_i > 0$ with
$\omega_r = 0$ and 
therefore is a non-propagating or {\it absolute} instability
(see for example Lifshitz and Pitaevskii 1981).  We assume axisymmetric
stability.  (If this is not the case, instability appears in the $m \geq 1$
waves giving growth independent of the counter-streaming.)

Figure 6a shows the one-armed ($m=1$) star and gas mode lines for a high mass
fractions of gas and $c_s/\sigma_r=0.316$.
For the near-crossing shown in Figure 6a, the mode lines `attract' in the
vicinity of the circle to give
instability as shown in Figures 6b and 6c.  (Conversely, the mode lines
`repel' in the case
where 
the crossing would be stable.)  Figure 7 shows the two-armed ($m=2$) mode
lines and instability also
for a high
mass fraction of gas.  Figure 8 shows the dependence of the
maximum 
growth rate of the one-armed mode on $\xi_g$, $Q_*$, and $c_s/\sigma_r$.

Consider the consequence of the wave growth for the cases
shown in Figures 6 ($m=1$) and 7 ($m=2$)..  We comment first on
the case of Figure 7 which
is similar to the case discussed in the previous sub-section.  Because
$\omega_r < 0$,
$f(\overline k_r) <0$ and 
$d \overline k_r/dt <0$ over the entire unstable range
of $\overline k_r$.  Over 
most of this range $d r/dt >0$ for $\overline k_r >0$,
whereas $d r/dt <0$ for $\overline k_r <0$.  Thus an unstable trailing
spiral wave
moves radially outward, whereas an unstable leading spiral wave moves
inward.  The
maximum 
amplification from equation (10) is about $\exp[2.4(k_{crit}r_o)]$ for
the trailing waves and $\exp[1.7(k_{crit}r_o)]$ for the leading waves.
Thus, for $k_{crit} = 2\pi/10kpc$ and $r_o = 10kpc$, the amplification factor
for the trailing waves is about
$~4.2\times 10^6$.

The case of Figure 6b for $m=1$ and $\xi_g=0.25$ is different
because $\omega(\overline k_r)$ changes sign
at $\overline k_0$
within 
the unstable range of $\overline k_r$.  Recall that for a flat rotation
curve, $r = r_o(\Omega/\omega_r)f(\overline k_r)$, where $r_o$
is the initial radius of the wave packet.  Thus, for $\overline k_r <
\overline k_0$ in the unstable range and $\overline k_r>0$, we
must have $\omega_r = const.>0$, so
that $f(\overline k_r) > 0$ and $d\overline k_r/dt >0$.
However, as 
$\overline k_r$ increases, $f(\overline k_r)$ eventually decreases
and 
approaches zero so that $r \rightarrow 0$;  that is, the wave packet moves
towards the center of the galaxy.  On the other hand, for $\overline k_r
>\overline k_0$
in the unstable range, we must
have $\omega_r=const. <0$ so that $f(\overline k_r)<0$,
and $d \overline k_r/dt <0$.
In this limit, $\overline k_r$ decreases and $f(\overline k_r)$
eventually increases and approaches
zero at $\overline k_0$ so that again $r \rightarrow 0$.
For both $\overline k_r > \overline k_0$ and $<\overline k_0$,
the wave amplification is even larger than
than that for the above-mentioned $m=2$
instability for the same $\xi_g$.  Thus,
it is likely that $A=A_{max}$.
The behavior of the trailing
waves, $\overline k_r<0$, is different.  For both $\overline k_r <
-\overline k_0$
and $> -\overline k_0,~$
$\overline k_r$ moves away from $-\overline k_0$
and the wave packet moves from its initial
radius to larger distances.  The wave
amplification from equation (10) in this case is even larger
than that for the trailing waves
because $|\Omega_o/\omega_r|\gg 1$ in equation (10).
When the wave-packet reaches and passes
the right or left-hand limit of the unstable
range of $\overline k_r$ for $\overline k_r <0$,
the packet continues to propagate adiabatically (without growth) away from
$\overline k_0$
with part of the wave energy in the upper frequency ($\omega_r$) branch and
part in the
lower branch.  For example, $\overline k_r$ larger than the right-hand
limit the wave on
the  lower frequency branch moves towards the center of the galaxy whereas
the wave on the
upper branch moves to larger radii.

For the case of Figure 6c for $m=1$ and $\xi_g=0.15$, $~\omega_r$ is positive
throughout the unstable range of $\overline k_r$.  The maximum amplification
factor for the trailing waves ($\overline k_r > 0$) is
about $\exp[3.5(k_{crit}r_o)]$,
whereas for the leading waves ($\overline k_r < 0$) it is
about $\exp[8.67(k_{crit}r_o)]$.
Both factors are very large for $k_{crit}r_o \sim 2\pi$, but that for the
leading
waves is strongly dominant.

Consider the $m=1$ wave growth for large mass-fractions of gas, say, $\xi_g =
0.3 - 0.7$.  Stability of the $m=0$ Toomre mode requires larger values of
$Q_*$ and/or $c_s/\sigma_r$ than considered above (see Fig. 5), and here we
consider $Q_*=1.8$ and $c_s/\sigma_r=0.5$.  For $\xi_g=0.3$,
the unstable waves occur for $\overline k_r = 0.72 -1.46$,
the maximum growth rate is $\omega_i/\Omega =0.17$, $\omega_r/\Omega$
varies from $0.092$ to $0.096$.
For $\xi_g =0.5$, the unstable range
is $\overline k_r = 0.63 - 2.0$, the maximum growth rate is
$\omega_i/\Omega =0.265$,
and $\omega_r/\Omega$ varies from zero to
$-0.042$.  For $\xi_g=0.7$, the unstable range
is $\overline k_r = 0.64 -2.56$,
the maximum growth rate is $\omega_i/\Omega = 0.29$, and $\omega_r/\Omega$
varies from $-0.13$ to $-0.2$.  Although the growth rate is somewhat larger
for $\xi_g = 0.7$, the wave amplification is the maximum possible ($A_{max}$)
for the $\xi_g = 0.5$ case where the wave frequency $\omega_r(\overline k_r)$
goes to zero in the unstable range of $\overline k_r$ (see equation 10).

In Section 5 we discuss non-linear processes which may lead to
saturation of the
wave growth.

\subsection{Co-Rotating Stars/Low-Mass Counter-Rotating Stars}

The dispersion relation for a two component galaxy consisting of
co-rotating ($+\Omega$) stars and a mass fraction $\xi_*\leq {1\over 2}$
of counter-rotating stars ($-\Omega$) is
$$0=\epsilon(\omega,k_r) \equiv 1+(1-\xi_*){\cal P}_*(s,k_r)+\xi_*{\cal
P}_*(s+w,k_r)~,
\eqno(12)$$
where $w \equiv 2 m \Omega/\kappa$ and ${\cal P}_*$ is defined below
equation (2).
The co-rotating star modes are given by $0 = 1+(1-\xi_*){\cal P}_*(s,k_r)$,
while the counter-rotating star modes are given by $0 = 1+\xi_*{\cal
P}_*(s+w,k_r)$.

Figure 9a shows the behavior of the two-armed ($m=2$) modes for $\xi_* \ll
1$ and
$Q_* = 1.4$ for both components.  The circled mode crossings in this
figure satisfy equation (7) suggesting instability.  The instability is
confirmed
by Figure 9b which shows the complex $\omega$ as a function
of $\overline k_r$
obtained from equation (12) for $\xi_*=0.05$ and $Q_*=1.4$ for
both components. From equation (10), the largest amplification factor
is about $\exp[0.18(k_{crit}r_o)]$ for the wave with $\omega_r <0$.  For
$k_{crit}r_o \sim 2\pi$, this amplification is insignificant.  For the same
conditions, the three-armed spiral waves are stable. The one-armed waves
are unstable for $\overline k_r = 0.935$ to $1.5$, with a maximum growth
rate of $\omega_i/\Omega = 0.079$ at $\overline k_r=1.25$ where
$\omega_r/\Omega=0.27$,
and with maximum amplification factor
$\exp[0.13(k_{crit}r_o)]$, which is insignificant
for $k_{crit}r_o \sim 2\pi$.

\subsection{Comparable Mass Co-/Counter-Rotating Stars}

Figure 10a shows the behavior of the two-armed ($m=2$) modes for $\xi_*
={1\over 2}$ and
$Q_* =1.4$ for both components.  Again the circled mode crossings in this
figure satisfy
equation (7) suggesting instability.  The instability is confirmed by
Figure 10b which
shows $\omega_r$ and $\omega_i$ as a function of $\overline k_r$
obtained from equation (12). From
equation (10), the maximum amplification
factor is about $\exp[0.58(k_{crit}r_o)]$.
 Note that for $\xi_* = {1\over 2}$, the waves
are completely symmetric under $\phi \rightarrow -\phi$:
the co-rotating wave ($\omega_+
=\omega_r+i\omega_i~,~\omega_r > 0$) is matched by an equivalent
counter-rotating wave
($\omega_- = -\omega_r+i\omega_i$) with the same growth rate.
Figure 11 shows the dependence of the maximum growth rate on $Q_*$.

The three-armed waves are
stable for $\xi_*={1\over 2}$ and $Q_*=1.4$ for both components.
However, the one-armed waves are strongly unstable for the same conditions.
The mode lines shown in
Figure 12a do not cross but `attract' (as noted earlier) near the circled
point to
give instability as shown in Figure 12b.  The exact
symmetry $\phi \rightarrow
-\phi$ for $\xi_* = {1 \over 2}$
makes it evident that the co- and counter-rotating
mode lines in Figure 12a must
merge at
$s= -\Omega/\kappa$ to give $\omega_r = 0$ (independent
of $\Omega/\kappa$) in
the range of $\overline k_r$ where $\omega_i >0$.  Thus the $m=1$
instability  in
this case is an {\it absolute} instability.  In this respect it
is similar to the $m=0$ instability.  Trailing and leading wave perturbations
can be superposed to give a standing wave in place of equation (1), $\delta
\Phi=
C \cos(\int^r dr'k_r(r'))\sin(\phi)\exp(\omega_it)$.  Figure 13 shows the
dependence of the maximum growth rate on $Q_*$ for $\xi_*={1\over 2}$.

For $\xi_* = {1\over 2}$, but unequal velocity spreads of the components,
the symmetry
$\phi \rightarrow -\phi$ is spoiled.  For example, the counter-rotating stars
may be younger with a smaller velocity spread.  For $\sigma_r^- = 0.316
\sigma_r^+$
($Q_*^-=0.316Q_*^+$), we find that stability of the $m=0$ Toomre mode
requires $Q_*^+
>1.86$, while the $m=1$ mode is unstable for $Q_*^+ <2.85$.  At
>$Q_*^+=1.86$, the
$m=1$ mode has
$\omega_i/\Omega \approx 0.34$ and $\omega_r/\Omega \approx = -0.2$.  In this
case the $m=1$ instability is {\it convective}.

For $\xi_* < {1\over 2}$, the one-armed waves still have large growth rates
if say
$\xi_* > 0.05$.  This is shown in Figure 14a.  However, $\omega_r$ takes on
positive
values due to the stronger `pull' of the co-rotating stars.  With $\omega_r
>0$,
$\overline k_r$ increases through the unstable range (see equation 9) for
both trailing
($\overline k_r > 0$) and leading ($\overline k_r < 0$) waves.  Thus
the instability is {\it convective}.  There is a weak dependence
of $\omega_r/\Omega$ on $\overline k_r$ which is such that the centroid of
a wave packet moves outward and then inward over a small range of $r$ as
$\overline k_r$ increases through the unstable range.
The maximum amplification factor from equation (10) is shown in
Figure 14b.

For $\xi_* >{1\over 2}$, the solutions are given by those for $\xi_* <
{1\over 2}$
by the replacements $\xi_* \rightarrow 1-\xi_*$ and $(\omega_r,~\omega_i)
\rightarrow
(-\omega_r,~\omega_i)$, provided that the velocity dispersions (the $Q_*'s$)
of the two components are the same.

\subsection{Co-Rotating Stars/Counter-Rotating Gas and Stars}

The dispersion relation for a three component galaxy consisting of
co-rotating stars and a mass fraction $\xi_g$ of
counter-rotating gas and a fraction $\xi_*$ of counter-rotating stars is
$$0=\epsilon(\omega,k_r) \equiv 1+(1-\xi_g-\xi_*){\cal
P}_*(s,k_r)+\xi_g{\cal P}_g(s+w,k_r)
$$ $$+\xi_*{\cal P}_*(s+w,k_r)~,
\eqno(12)$$
where $w \equiv 2 m \Omega/\kappa$ and ${\cal P}_*$ is defined below
equation (2).
The co-rotating star modes are given by $0 \approx 1+(1-\xi_g-\xi_*){\cal
P}_*(s,k_r)$,
while the counter-rotating gas modes are given by $0 = 1+\xi_g{\cal
P}_g(s+w,k_r)$
and the counter-rotating star modes are given by
$0 = 1+\xi_*{\cal P}_*(s+w,k_r)$.

Figure 15 shows the real and imaginary parts of $\omega$ as a function of
$\overline k_r$
for one-armed waves for a sample case where
$\xi_g = \xi_* = 0.1$, with $Q_* =1.4$ for both star components, and
$c_s/\sigma_r = 0.316$ for the gas.  The instability arises mainly from the
interaction
of the co-rotating stars and the counter-rotating gas due to the lower
velocity dispersion of the gas.  The maximum amplification factor for the
trailing
waves ($\overline k_r >0$) is about $\exp[1.23(k_{crit}r_o)]$, whereas for
the leading
waves ($\overline k_r <0$) it is about $\exp[1.7(k_{crit}r_o)]$.
For $k_{crit}r_o = 2\pi$, these factors
are $2.3\times 10^3$ and $4.3 \times 10^4$, respectively.

Figure 16 shows the
dependence of the
one-armed ($m=1$) wave instability on $\xi_g$ for three values of
$\xi_*$.
In this 
figure, $Q_*=1.6$ for both
stellar components, and $c_s/\sigma_r=0.5$. From the discussion
of Sec. 3.3, it
is clear that the largest amplification occurs when $\omega_r=0$ within
the unstable 
$\overline k_r$ range.  Therefore, it follows from this Figure 16b
that the largest amplification occurs for $\xi_g +\xi_* \approx {1\over 2}$.
Under this condition the amplification is largest for the leading waves
($\overline k_r < 0$).  The growth rates and amplification factors are
smaller for the $m=2$ waves.

\section{POSITIVE/NEGATIVE MODE ENERGY OF UNSTABLE WAVES}

The energy-density of an electrostatic wave in a homogeneous plasma is given
by the well-known expression ${\cal E}_w = \omega_r
(\partial \epsilon/\partial \omega_r)|\delta {\bf E}|^2/(8 \pi)$, where
$\epsilon$ is the dielectric function, $\delta {\bf E} = -{\bf\nabla}\delta
\Phi_e$,
and $\nabla^2 \delta \Phi_e = - 4\pi \delta \rho_e$ (Coppi, Rosenbluth, \&
Sudan 1969).
The corresponding expression for a self-gravitating medium differs
by a minus sign and is $ {\cal E}_w =
-\omega_r(\partial \epsilon/\partial \omega_r)|\delta {\bf g}|^2/$ $(8\pi
G)$, where
$\delta {\bf g} = -{\bf \nabla} \delta \Phi$ and $\nabla^2 \delta \Phi = +
4\pi G \delta \rho$.
The generalization to tightly wrapped waves in a flat galaxy gives the
energy (per unit
area of the disk)
$$ {\cal E}_w = - \omega_r(\partial \epsilon/\partial \omega_r)
\pi G |\delta \Sigma|^2/|k_r|~,
\eqno(13)$$
where we have used $\delta \Phi = -2 \pi G \delta \Sigma/|k_r|$, and
$\delta \Sigma$
is the total surface mass-density perturbation.  An equation equivalent to
(13) is
derived by Shu (1992).

For a galaxy
consisting of co- and counter-rotating components (now denoted for
generality by $+$ and $-$), we have $\epsilon = 1 + {\cal P}^+ +{\cal
P}^-$.  Thus,
the wave energy-density is the sum of the energy densities associated with
the modes in the co-
and counter-rotating components,
$${\cal E}_w = {\cal E}^+_w +{\cal E}^-_w,$$
$${\cal E}_w^\pm = -\omega_r(\partial {\cal P}^\pm/\partial\omega_r)\pi
G|\delta \Sigma|^2/|k_r|~.
\eqno(14)$$
If the signs of ${\cal E}_w^+ $ and ${\cal E}_w^-$ are different, we have the
necessary condition for the well-known
instability of interacting positive and negative energy modes (Coppi et al.
1969). The
negative energy mode can grow by feeding energy into the positive energy
mode.  A plasma
example is the two-stream instability which can occur when a beam of
charged particles
passes with velocity $v_b$ through a background plasma.
In this case the energies associated with the modes
in the beam
and the background plasma have opposite signs.
The mechanism of this instability can be
understood by considering a localized
excess or clump of say positive charge in the
beam which will induce a negative clump in the background.
The relative motion of
the clumps leads to an electrostatic attraction
between them, a slowing down of the beam clump,
a speeding up of the background clump,
and a growth of the electric field energy.
The clump size must be sufficiently large to give instability,
larger than about $\pi v_b/\omega_p$ for the case of equal beam and plasma
densities, where $\omega_p$ is the plasma frequency.
In a {\it homogeneous} self-gravitating
system involving a beam and a background,
a density excess in the beam will
induce an excess in the background, and it would
appear that a similar two-stream
instability would occur. However, for most distribution
functions, the large clump size
needed for the two-stream instability ($>\pi v_b
/\omega_J$, where $\omega_J=(4\pi G \rho)^{1\over2}$ is the Jeans angular
frequency)
results instead in the Jeans instability being dominant
(Lynden-Bell 1967, Araki 1987).  The Jeans instability occurs for clump
sizes larger than about $\pi \Delta v/\omega_J$, where $\Delta v$ is the
velocity spread. In contrast,
in a disk galaxy the velocity spreads of the $\pm \Omega$
components {\it and } the disk rotation [$(\kappa^+)^2 = (\kappa^-)^2 >0$]
provide stability to the Jeans instability ($m=0$) (Toomre 1964).

{}From equations (2) and (3), the condition
for ${\cal E}^+_w{\cal E}^-_w < 0$ is
simply $s_r(s_r+w)<0$ or $-\Omega <\omega_r/m < \Omega$ which
is equation (7).
Recall that $\omega_r/m$ is the pattern or angular phase velocity of
the wave.  For the two-stream instability in a beam-plasma system, the phase
velocity of the unstable waves is also intermediate between velocity of the
beam
and that of the background.

The axial angular momentum of the wave (per unit area of the disk) can be
written
as ${\cal J}_w = {\cal J}_w^+ + {\cal J}_w^-$, where ${\cal J}_w^\pm = -
m(\partial
{\cal P}^\pm/\partial \omega_r)\pi G|\delta \Sigma|^2/|k_r|~$ (Coppi et al.
1969),
so that ${\cal E}_w = (\omega_r/m){\cal J}_w$.
Thus, the wave angular momenta associated with the modes in the co- and
counter-
rotating components have opposite signs under the same condition that the
mode
energies have opposite signs.  Under this condition, the angular momentum of
the mode in the co-rotating component is negative, whereas that in the
counter-rotating component is positive.  Thus, the total angular momentum
of the mode plus matter of the $+\Omega$ component is reduced, whereas
that in the $-\Omega$ component is increased (but decreased in magnitude).
The
situation is analogous for the beam-plasma instability where the linear
momentum of the beam mode is negative and that of the background mode is
positive.

\section{NON-LINEAR EFFECTS AND GAS ACCRETION}

The non-axisymmetric instabilities discussed here are of interest because
they could be important for the inward radial
transport or accretion of counter-rotating ($-\Omega$) matter
deposited or accreted at large $r$ onto an existing flat galaxy consisting
mainly of co-rotating ($+\Omega$) stars.
We consider first the case where the
counter-rotating matter is gas.

The linear wave growth is determined mainly by three parameters, the Toomre
parameter $Q_*$ for the stars, the ratio of the sound speed in the gas to
the radial velocity spread of the stars $c_s/\sigma_r$, and the mass fraction
of the counter-rotating gas $\xi_g$.  In general these quantities will vary
with $r$;  the theory is still valid as long as the variation is small on a
scale $1/k_r$.  For a given $\xi_g$, the wave growth rate decreases
monotonically
with increasing $Q_*$ and $c_s/\sigma_r$, as shown
for example by Figure 8 for one-armed waves.
On the other hand, with $Q_*$ and $c_s/\sigma_r$ fixed, the wave growth
is zero
for $\xi_g$ below a threshold value and increases strongly as $\xi_g$
increases
above this value as shown by Figure 8a.

Suppose that $Q_*,~c_s/\sigma_r$, and
$\xi_g$ are such as to give a large
growth rate, say, $\omega_i/\Omega > 0.05$,
and an appreciable wave amplification factor, $ A \gg 1$.  After several
rotation
periods of the matter at a radius $r$, the wave amplitude can grow to a level
where non-linear effects become
important ($|\delta \Phi| \sim (0.01 - 0.05) (\Omega
r)^2$).  One non-linear effect is the scattering of stars into less
circular orbits
which acts to increase the radial velocity spread and therefore $Q_*$.
In turn,
the increase of $Q_*$ acts to reduce the growth rates.
However, a finite level of non-axisymmetric waves may remain excited due to
reduced but non-zero wave growth.
These waves can give an effective
viscosity $\nu_{e}$ which causes inward transport
or accretion
of the gas and outward transport of the gas's angular momentum.
The gas can remain in approximately circular motion
by radiating the energy excess which
results from its inward motion.  This gives an
accretion luminosity (per unit
area of the disk)  $ \Sigma_g \Omega^2 r
|\overline v_{gr}|$, assuming a flat rotation curve,
where $\overline v_{gr}<0$ is the average accretion speed of
the gas.  A rough estimation using quasi-linear theory gives
$\nu_{e} \sim \Omega r^2 |k_r r|^3 |\delta \Phi/(\Omega r)^2|^2$
and ${\overline v_{gr}} \sim -\nu_e/r$, where $\delta \Phi$
is the residual wave level.  Relevant values for accretion from
say $r=20$ kpc in $3\times 10^9$ yrs. are $|v_{gr}| > 5$ km/s.

A second non-linear effect is the
steepening of the spiral wave in the gas which can lead to the
formation of a oblique shock as indicated in Figure 17 for the case
of a one-armed leading wave.  (The amplification factor for the leading
wave is generally larger than that for the trailing wave.)  The gas
velocity parallel to
the shock is unchanged across the shock.  Up stream of the shock, the
normal component
of the velocity $v_{gn} = \theta \Omega r$ is larger than the ambient sounds
speed $c_s^o$ in order to have a shock.  
Here, $\theta =|k_r r|^{-1} \ll 1$ is
the pitch angle of the spiral arm.  That is, the normal Mach number
$M_n=v_{gn}/c_s^o >1$.  We assume that $M_n$ is not much larger than unity so
that the shock is not very strong.
Downstream from the shock, the normal velocity of the gas is less
than the post-shock sound speed.  As a result, the gas acquires
an inward radial velocity
$v_{gr} = -\beta \Omega r$, where $ \beta = {{2}\over{\gamma+1}}
(1-1/M_n^2) \theta~$ and $\gamma$ is the adiabatic index.  This shock
deflection could be important for the inward radial transport of
counter-rotating
gas.  The angular momentum lost by the gas may in be transported outward
by the spiral wave.
Also in this case, the gas
can remain in approximately circular motion by radiating the energy excess
arising from accretion.
Across the disk, the azimuthally averaged accretion speed $\overline v_{gr}$
has a strong non-linear dependence on $\xi_g$ because at small $\xi_g$ there
is no instability and no transport, while
for $\xi_g \approx {1\over 2}$ there is maximum wave amplification
(Sec. 3.3).
As a result, the gas transport equation
$\partial \Sigma_g/\partial t + (1/r)\partial [r \Sigma_g {\overline
v}_{gr}(\Sigma_g)]/
\partial r = 0$ (neglecting star formation) is strongly non-linear.
Its solutions may involve inward propagating
soliton-like perturbations in $\Sigma_g$.  This may be the cause of
ring-like features
observed in some counter-rotating galaxies (for example, NGC 7217, Buta et
al. 1995,
and NGC 4138, Jore et al. 1996).

The growth of spiral arms with significant self-gravity
may be important in inducing collisions and agglomeration of gas clouds
in the
arms (see for example Roberts, Lowe, \& Adler 1990) thereby enhancing
the rate of formation of counter-rotating stars.  Newly formed
counter-rotating
stars should have a smaller velocity spread than the older co-rotating stars
which have undergone stochastic heating (BT, p. 484).

If the counter-rotating matter consists of stars, the $m=1$ instability
discussed
in Sec. 3.2 may be important.  From Figure 14, the instability is
significant for
a mass fraction of counter-rotating stars $\xi_*$ larger than $0.05$ and is
strongest for $\xi_* = {1\over 2}$ if $Q_* = 1.4$ for both components.  For
$\xi_*$ increasing from ${1\over 2}$, the growth rate decreases.  The
large growth rates and amplification factors of Figure 14 we expect to
exist only transiently because the resulting waves would rapidly scatter
stars
into less circular orbits
increasing $Q_*$ and decreasing $\omega_i$.
However, a finite level of waves may remain excited due to much smaller
growth rates.  Consider now the possibility of radial inflow of the stars.
A counter-rotating star at a radius $r$ with angular
momentum ${ l}^-_z < 0$ can lose angular momentum $\delta { l}_z^->0$ to
the spiral wave in a given period of time so that the magnitude of $ l_z^-$
decreases.  At the same time, a co-rotating star also at $r$ with angular
momentum $ l_z^+ > 0$ can lose $\delta l_z^+ <0 $ so that the magnitude of
$l_z^+$ also decreases.  The resulting smaller
values of $|l_z^\pm|$ correspond to smaller
mean radii.  In order for a co- or counter-rotating star to stay in
approximately circular motion, its energy must decrease by an amount
$\delta e^\pm = \Omega \delta l_z^\pm < 0$, assuming a flat rotation curve.
However, the ratio of the energy to the angular momentum of the spiral
wave ${\cal E}_w/{\cal J}_w = \omega_r/m = \omega_r \ll \Omega$ from
Sec. 4 and Figure 14.  That is, the energy loss of the stars to the
spiral wave is too small to allow inward radial motion on approximately
circular orbits.  Even a small fractional inward motion of the stars
will increase $Q_*$ sufficiently to stabilize the spiral wave modes.

The stability of a galaxy with both counter-rotating stars and
gas (Sec. 3.6) appears qualitatively similar
to the results found with only counter-rotating gas.  The largest
wave amplifications
are for the $m=1$ leading spiral waves and for $\xi_*+\xi_g \approx
{1\over2}$.
The above discussion of gas transport is also pertinent to this case.

\section{CONCLUSIONS}

The two-stream instability - well-known in plasma physics - has
been discussed by a number of authors for
counter-streaming self-gravitating systems.  However, a somewhat
muddled picture has emerged for the role of this instability in
galaxies with counter-rotating components including the notion
that the instability does not occur due to the difference between
the universal attraction of masses and the repulsion/attraction of
like/unlike charges.  Indeed, stability analysis of a
homogeneous self-gravitating system of
counter-streaming particles
(following Jeans' neglect of the background potential),
shows that the two-stream instability
is dominated by the
Jeans instability if the distribution
functions are Maxwellian (Lynden-Bell 1967,
Araki 1987).  However, the homogeneous
system results are {\it not} relevant
to counter-stream matter in disk
galaxies where the Jeans instability ($m=0$)
is stabilized by both the velocity
spreads of the two components {\it and}
the disk rotation ($\kappa^2 > 0$
for both components) (Toomre 1964).  A
study by Araki (1987) of the stability of counter-streaming,
rigidly-rotating,
finite-radius Kalnajs disks does in fact show an $m=1$ two-stream
instability,
but the physical nature of the instability is not elucidated.
Simulation studies of different galaxy
models by Zang and Hohl (1978), Sellwood
and Merritt (1994), and Howard et al. (1996) all show clear evidence of a
strong $m=1$ two-stream instability.

In this work, we first briefly summarize known results for small-amplitude,
tightly-wrapped spiral waves in a single component stellar disk and
show (in Sec. 2) that the higher order modes of the dispersion relation have
been overlooked in standard treatments (BT, Palmer 1994).  The
dispersion relation gives the dependence of the wave
frequency ($\omega$) on the radial wavenumber ($k_r$), and the
omitted curves are labeled $l=2,3,..$ in Figure 2.  These higher order
modes are the analogues of the Bernstein modes in a collisionless plasma
which propagate across a uniform magnetic field.  Their existence
leads to a richer set of Linblad resonances which occur at pattern
speeds $\Omega_p \equiv \omega/m = \Omega \pm l\kappa/m$ with $m=1,2,..$
{\it and} $l=1,2,..$.

We assume tightly wrapped spiral waves, $|k_r| r \gg 1$, so
that the WKB approximation holds, but it is not known how strong this
inequality must be.  On the other hand, for effective
swing amplification involving radial reflection of waves, which
is thought to drive spiral waves in galaxies with only co-rotating matter,
one needs $|k_r|r <  3$ (BT, p. 376).  Here, the two-stream wave
amplification is sufficiently large
so as to be important without wave reflection.
Although selection effects may be
involved, observed counter-rotating disk galaxies
are early-type spirals or S0's
which do not show prominent open spiral structure.

We go on to discuss spiral waves in flat galaxies with counter-streaming
where
there is a co-rotating stellar component with angular velocity $+\Omega$
and a
counter-rotating component(s) of gas and/or stars with angular
velocity $-\Omega$.  We then plot on the same graph the
$\omega = \omega(k_r)$ lines - referred to as mode lines - for the
co-rotating stellar
component (neglecting the counter-component) and the $\omega = \omega(k_r)$
lines
for the counter-component (neglecting the co-component).  At crossing points
of these mode lines in the $(k_r,\omega)$ plane, a strong
resonant interaction can occur between co- and counter-rotating components.
Direct solution of the dispersion relations (Sec. 3) shows that
the crossings are unstable if the pattern angular
speed of the spiral wave $\omega/m$ is
such that $-\Omega < \omega/m < \Omega$.  The `mode-crossings' and
near `mode-crossings' (Sec. 3.3) satisfying this condition give
the two-stream instability in a counter-rotating galaxy.  They arise from the
fact that the mode energies for the co- and counter-rotating components have
opposite signs for $-\Omega < \omega/m < \Omega$ (Sec. 4).
This is analogous to the two-stream instability in a beam-plasma system
where the mode energies for the beam and the background have opposite signs
for a wave phase velocity intermediate between that of the beam and that of
the background.

Instability of a spiral wave may or may not be significant because the
waves are in most cases {\it convective} in the sense that both the
position ($r$) and the wavenumber ($k_r$) of an unstable wave packet
evolve with time (Sec. 3.2).  More important than the local growth rate
($\omega_i$) is the amplification factor of the advected wave $A$ (equation
10)  which is the maximum factor by which an
unstable packet can grow before being convected out of the unstable range
of $k_r$.  In most cases, we find that the amplification
factors are largest for one-armed $m=1$ leading spiral waves (with respect
to the co-rotating stars).
However, for larger stellar velocity
spreads (larger $Q_*$), the two-armed $m=2$
wave may be unstable, with the largest amplification
for the trailing wave, while the $m=1$ wave is
stable.   The
growth rates and amplification factors increase as the mass-fraction of
counter-rotating gas $\xi_g$ or of stars $\xi_*$ increases if these
fractions are less than $\approx {1\over 2}$.
On the other hand, both $\omega_i$ and
$A$ decrease as the velocity spreads of the stars and/or gas increase.

For the case of only counter-rotating gas of
mass fraction $\xi_g < \approx {1\over 2}$,
the largest amplification factors are for the $m=1$ leading spiral waves
(Sec. 3.3).
For the case of only counter-rotating
stars of mass-fraction $\xi_* < \approx{1\over 2}$,
the largest $\omega_i$ and $A$ values
occur also for the $m=1$ leading spiral
waves (Sec. 3.5).  If the velocity spreads in the
two stellar components are equal, then for $\xi_* = {1\over 2}$, the $m=1$
instability is an {\it absolute} or non-propagating instability, the real
part of the wave frequency ($\omega_r$) is zero, and the growth rate
has its maximum value.  For $\xi_* > {1\over 2}$, the instability is again
convective, the amplification factors are largest for the trailing spiral
waves, and $\omega_i$ and $A$ decrease with increasing $\xi_*$.
These results are in qualitative accord
with the theoretical results of Araki (1987) for
counter-rotating
stellar disks and with the simulation results of Sellwood and
Merritt (1994), where $\xi_* = {1\over 2}$ in both studies.

For a galaxy with both counter-rotating stars and gas, the largest
amplification is for the $m=1$ leading spiral waves, and
it occurs when the counter-rotating
mass fraction is $\xi_*+\xi_g \approx {1\over2}$ (Sec. 3.6).

Possible non-linear effects which act to limit the wave growth and
amplification are discussed in Sec. 5.  The effects include scattering
of star orbits by the wave which increases $Q_*$, and heating of the gas.
A residual level of spiral waves may remain excited which give an
effective viscosity for the gas causing its accretion.  Also, a leading
$m=1$ spiral shock wave may form in the gas causing its accretion.  
However, from the considerations of Sec. 5, it appears unlikely that
the spiral
waves cause accretion of counter-rotating stars.

The gas viscosity (and thus accretion rate) due to the two-stream
waves is plausibly
an increasing  function of the spiral wave amplification factor.  With no
amplification there are no waves and no viscosity.  Thus, the fact that the
largest amplification occurs for a mass-fraction of counter-rotating
matter $\xi_*+\xi_g \approx {1\over 2}$ may be pertinent to the
counter-rotating
galaxy NGC-4550 (Rubin et al. 1992) which is remarkably symmetric between
co- and counter-rotating stars, $\xi_* \approx {1\over 2}$ (Rix et al. 1992).
A schematic scenario is that initially counter-rotating gas is supplied
at large $r$ and has a fastest accretion rate for $\xi_g={1\over2}$.
At later times, after counter-rotating stars have formed with mass
fraction $\xi_*$, the fastest accretion rate
is for $\xi_g = {1\over 2} -\xi_*$.
At even later times, the gas accretion ceases
when $\xi_* \approx {1\over 2}$.

Accretion of gas due to two-stream waves is strongly nonlinear when
$\xi_g$ is large and this may lead to
a `pile up' of gas into rings at one or
more radial distances.  Prominent rings are seen in NGC 3593, NGC
4138, and NGC 7217.  Buta et al.(1995) have suggested that the locations
of the three rings in NGC 7217 correspond to different Linblad resonances.
At sufficiently high counter-rotating gas densities counter-rotating
star formation may be triggered.

\acknowledgments{

We thank G. Contopoulos, H.H. Fleischmann, B.R. Kusse,
W.I. Newman, J.A. Sellwood,
and A.R. Thakar for valuable
discussions.  This work was supported
in part by NSF grant AST-9320068.}

\newpage

\figcaption{Sketch of the counter rotating disk geometry. }

\figcaption{ The figure shows the frequency
[$s \equiv (\omega - m\Omega)/\kappa$] - wavenumber ($k_r$)
dependence of the tightly wrapped spiral wave modes in a disk
of stars.  Here, $\Omega$ is the angular velocity of the stars, $\kappa$ is
their
epicyclic frequency, $k_{crit}$ is the critical
wavenumber, and $l$ labels different branches
of the dispersion relation.  The higher branches ($l \ge 2$) are analogues
of the
Bernstein modes in a magnetized plasma as discussed in the text.}

\figcaption{The figure shows the different star
and gas two-armed ($m=2$) modes in a galaxy of co-rotating
stars with $Q_* = 1.4$ and a small mass
fraction of counter-rotating gas ($\xi_g \ll 1$).
The curves
extend to negative $k_r$ as even functions.
At the circled points where the star and gas modes cross
there is a strong resonant interaction which may lead to instability.
Here, only the point B is unstable.   The detailed behavior of
the mode crossings is shown in Figure 4.}

\figcaption{ The figure shows the behavior of the star/gas
mode crossings for two armed waves ($m=2$) and a small mass
fraction of gas, $\xi_g = 0.01$.  Also, $Q_*=1.4$, $c_s/\sigma_r = 1$, and
we have assumed a flat rotation
curve so that $\kappa = \sqrt2 \Omega$.
{\bf ~~a} is for an unstable case (point B of Figure 3), and
{\bf b} for a stable case (point C of Figure 3).}

\figcaption{ The figure shows the
dependence of the axisymmetric ($m=0$)
Toomre (1964) stability threshold on the sound
speed in the gas ($c_s$) obtained from equation (9).
Here, $\sigma_r$ is the radial velocity
dispersion of the stars, $\xi_g$ is the mass fraction of the gas, and
$Q_*$ is the Toomre stability parameter for the stars defined
below equation (2).  We have assumed $\kappa = \sqrt2 \Omega$.  Note that
the dashed curve $\xi_g = 1$ is given by $c_s/\sigma_r = 0.9355/Q_*$,
which is the same as $c_s=\kappa/(2k_{crit})$ from equations (2) and (3).}

\figcaption{ The figure shows the nature of the
one-armed ($m=1$) modes in a galaxy of co-rotating stars
and an appreciable mass fractions of counter-
rotating gas.  For this figure, $Q_*=1.4$, $c_s/\sigma_r =0.316$, and
$\kappa = \sqrt2 \Omega$.  {\bf~a~} shows the star and gas mode lines.
The circle indicates the near crossing
which is unstable.  {\bf~b~} shows the real and imaginary parts of $\omega$
obtained from equation (4) for $\xi_g=0.25$, and {\bf c}
shows the same quantities for $\xi_g =0.15$.}

\figcaption{ The figure shows the nature of the
two-armed ($m=2$) modes in a galaxy of co-rotating stars
and an appreciable mass fraction ($\xi_g = 0.25$) of counter-
rotating gas.  For this figure, $Q_*=1.4$, $c_s/\sigma_r =0.316$, and
$\kappa = \sqrt2 \Omega$.  {\bf~a~} shows the star and gas mode lines.
The circles indicate unstable mode crossings.
{\bf~b~} shows the real and imaginary parts of $\omega$
obtained from equation (4).}

\figcaption{ The figure shows the dependences of the maximum
growth rate of the one-armed waves on $\xi_g$, $Q_*$, and $c_s/\sigma_r$.
The left-hand limit in {\bf b} at $c_s/\sigma_r =0.2$ is close to the
stability threshold of the axisymmetric instability (see Fig. 5).}

\figcaption{ The figure shows the nature of the two-armed
($m=2$) modes in a galaxy of co-rotating stars
($+\Omega$) and a small fraction ($\xi_*\ll 1$)
of counter-rotating stars ($-\Omega$).  {\bf~ a~} shows the mode lines
with the
unstable crossings circled.  {\bf~b~} shows the real and imaginary parts of
$\omega$
obtained from equation (12).  For both star components, $Q_*=1.4$, and
we have
taken $\kappa =\sqrt2 \Omega$.  The vertical arrows on the $\omega_i$ curves
indicate the associated $\omega_r$ curves.}

\figcaption{ The figure shows the nature of
the two-armed ($m=2$) modes in a galaxy of co-rotating
stars ($+\Omega$) and an equal mass ($\xi_* = {1\over 2}$) of
counter-rotating
stars ($-\Omega$).
{\bf~a~} shows the mode lines with the unstable crossing circled.
{\bf~b~} shows the real
and imaginary parts of the wave frequency $\omega$ as a
function of $\overline k_r$ obtained from equation (12).
For both star components, $Q_* = 1.4$.  Also, we have
taken $\kappa = \sqrt2 \Omega$.  The vertical arrows
on the $\omega_i$ curves point towards the associated $\omega_r$
curves. }

\figcaption{ The figure shows the $Q_*$ dependence of
the maximum growth rate of the two-armed waves for $\xi_*= {1\over 2}$.
The value of $k_r$ for the maximum growth is also indicated.
We have
taken $\kappa = \sqrt2 \Omega$.}

\figcaption{ The figure shows the nature of
the one-armed ($m=2$) modes in a galaxy of co-rotating
stars ($+\Omega$) and an equal mass ($\xi_* = {1\over 2}$) of
counter-rotating
stars ($-\Omega$).
{\bf~a~} shows the mode lines, and the circle indicates the
unstable near crossing of mode lines.
{\bf~b~} shows the real
and imaginary parts of the wave frequency $\omega$ as a
function of $\overline k_r$ obtained from equation (12).
For both star components, $Q_* = 1.4$.}

\figcaption{ The figure shows the $Q_*$ dependence of
the maximum growth rate of the one-armed waves for a galaxy
of equal mass co- and counter-rotating components, $\xi_*= {1\over 2}$.}

\figcaption{ The figure shows the
maximum growth rate and amplification factor
of the one-armed waves for a galaxy of
co-rotating stars and an appreciable mass-fraction $\xi_*$
of counter-rotating stars.}

\figcaption{The figure shows the nature of
the one armed unstable mode of galaxy of co-rotating stars
and a fraction $\xi_g=0.1$ of counter-rotating gas and
a fraction $\xi_* = 0.1$ of counter-rotating stars.  For this
figure, $Q_*=1.4$ for both star components,
$c_s/\sigma_r =0.316$, and $\kappa = \sqrt2\Omega$.}

\figcaption{ The figure shows
the dependence of the one-armed ($m=1$) spiral wave instability
on $\xi_g$ for three values of $\xi_*$.  {~~\bf a~} shows the
dependence of $(\omega_r)_m$ on $\xi_g$, where
$(\omega_r)_m$ is the
value of $\omega_r$ at the point
where $|\omega_r(\overline k_r)|$
is a minimum for $\overline k_r$
within the unstable range.  {~~\bf b~} shows the
dependence of the maximum
growth on $\xi_g$.
The figure assumes $Q_*=1.6$ for both
stellar components, $c_s/\sigma_r = 0.5$, and
$\kappa = \sqrt 2 \Omega$.  The largest wave amplification occurs
when $(\omega_r)_m = 0$, and panel {~\bf a~} shows that this happens
for $\xi_*+\xi_g \approx {1 \over 2}$.}

\figcaption{ Geometry of a one-armed leading spiral
shock wave in counter-rotating gas which leads to inward
radial motion of the gas.}

\end{document}